\documentclass[education,article,accept,moreauthors,pdftex]{Definitions/mdpi}

\usepackage[normalem]{ulem} 
\usepackage{amsfonts} 
\usepackage{wasysym} 
\usepackage{mathcomp} 
\usepackage{CJKutf8} 
\usepackage{pifont} 
\usepackage{bm} 
\usepackage{bbm} 
\graphicspath{{./Definitions/}} 

\makeatletter
\def\T@n@@nc@d@ngM@cr@M@d{}
\def\LY@n@@nc@d@ngM@cr@M@d{}
\makeatother

\let\orignewcommand\newcommand  
\let\newcommand\providecommand  
\usepackage{verse}
\let\newcommand\orignewcommand  

\newsavebox\foobox

\renewcommand{\dagger}{\mathchar"2279}
\renewcommand{\ddagger}{\mathchar"227A}

\newcommand{\mmathit}[1]{
  \ifthenelse{\equal{#1}{\ln}}{\mathit{ln}}{
    \ifthenelse{\equal{#1}{\max}}{\mathit{max}}{\mathit{#1}}
  }
}
\makeatother
\robustify{\footnote} 

\DeclareUnicodeCharacter{1E45}{\.{n}}
\DeclareUnicodeCharacter{1E41}{\.{m}}
\DeclareUnicodeCharacter{2003}{\quad}
\DeclareUnicodeCharacter{2009}{\thinspace}
\DeclareUnicodeCharacter{2002}{\enspace{}}
\DeclareUnicodeCharacter{2005}{\thinspace}
\DeclareUnicodeCharacter{0263}{\textipa{G}}
\DeclareUnicodeCharacter{A0}{~}
\DeclareUnicodeCharacter{2460}{\textcircled{\scriptsize{1}}}
\DeclareUnicodeCharacter{2461}{\textcircled{\scriptsize{2}}}
\DeclareUnicodeCharacter{2462}{\textcircled{\scriptsize{3}}}
\DeclareUnicodeCharacter{2463}{\textcircled{\scriptsize{4}}}
\DeclareUnicodeCharacter{2464}{\textcircled{\scriptsize{5}}}
\DeclareUnicodeCharacter{2465}{\textcircled{\scriptsize{6}}}
\DeclareUnicodeCharacter{2466}{\textcircled{\scriptsize{7}}}
\DeclareUnicodeCharacter{2467}{\textcircled{\scriptsize{8}}}
\DeclareUnicodeCharacter{2468}{\textcircled{\scriptsize{9}}}
\DeclareUnicodeCharacter{2070}{\textsuperscript{0}}
\DeclareUnicodeCharacter{2074}{\textsuperscript{4}}
\DeclareUnicodeCharacter{2075}{\textsuperscript{5}}
\DeclareUnicodeCharacter{2076}{\textsuperscript{6}}
\DeclareUnicodeCharacter{2077}{\textsuperscript{7}}
\DeclareUnicodeCharacter{2078}{\textsuperscript{8}}
\DeclareUnicodeCharacter{2079}{\textsuperscript{9}}
\DeclareUnicodeCharacter{02C2}{<}
\DeclareUnicodeCharacter{2033}{\relax\ifmmode '' \else $''$\fi}
\DeclareUnicodeCharacter{2034}{\relax\ifmmode ''' \else $'''$\fi}
\DeclareUnicodeCharacter{2026}{\relax\ifmmode … \else $\ldots$\fi}
\DeclareUnicodeCharacter{0229}{\c{e}}
\DeclareUnicodeCharacter{016F}{\r{u}}
\DeclareUnicodeCharacter{127}{\relax\ifmmode\rm\hbar\else $\rm\hbar$\fi}
\DeclareUnicodeCharacter{3AC}{\relax\ifmmode\acute{\alpha}\else $\acute{\alpha}$\fi}
\DeclareUnicodeCharacter{3AD}{\relax\ifmmode\acute{\varepsilon}\else $\acute{\varepsilon}$\fi}
\DeclareUnicodeCharacter{3AE}{\relax\ifmmode\acute{\eta}\else $\acute{\eta}$\fi}
\DeclareUnicodeCharacter{3AF}{\relax\ifmmode\acute{\iota}\else $\acute{\iota}$\fi}
\DeclareUnicodeCharacter{3CC}{\relax\ifmmode\acute{o}\else $\acute{o}$\fi}
\DeclareUnicodeCharacter{3CD}{\relax\ifmmode\acute{\upsilon}\else $\acute{\upsilon}$\fi}
\DeclareUnicodeCharacter{3CE}{\relax\ifmmode\acute{\omega}\else $\acute{\omega}$\fi}
\DeclareUnicodeCharacter{391}{A}
\DeclareUnicodeCharacter{392}{B}
\DeclareUnicodeCharacter{395}{E}
\DeclareUnicodeCharacter{396}{Z}
\DeclareUnicodeCharacter{397}{H}
\DeclareUnicodeCharacter{399}{I}
\DeclareUnicodeCharacter{39A}{K}
\DeclareUnicodeCharacter{39C}{M}
\DeclareUnicodeCharacter{39D}{N}
\DeclareUnicodeCharacter{39F}{O}
\DeclareUnicodeCharacter{3A1}{P}
\DeclareUnicodeCharacter{3A4}{T}
\DeclareUnicodeCharacter{3A7}{X}

\DeclareUnicodeCharacter{00B5}{\relax\ifmmode \mu{} \else $\mu$\fi}
\DeclareUnicodeCharacter{27E6}{\relax\ifmmode \llbracket \else $\llbracket$\fi}
\DeclareUnicodeCharacter{27E7}{\relax\ifmmode \rrbracket \else $\rrbracket$\fi}
\DeclareUnicodeCharacter{1D434}{\relax\ifmmode A \else $A$\fi}
\DeclareUnicodeCharacter{1D435}{\relax\ifmmode B \else $B$\fi}
\DeclareUnicodeCharacter{1D436}{\relax\ifmmode C \else $C$\fi}
\DeclareUnicodeCharacter{1D437}{\relax\ifmmode D \else $D$\fi}
\DeclareUnicodeCharacter{1D438}{\relax\ifmmode E \else $E$\fi}
\DeclareUnicodeCharacter{1D439}{\relax\ifmmode F \else $F$\fi}
\DeclareUnicodeCharacter{1D43A}{\relax\ifmmode G \else $G$\fi}
\DeclareUnicodeCharacter{1D43B}{\relax\ifmmode H \else $H$\fi}
\DeclareUnicodeCharacter{1D43C}{\relax\ifmmode I \else $I$\fi}
\DeclareUnicodeCharacter{1D43D}{\relax\ifmmode J \else $J$\fi}
\DeclareUnicodeCharacter{1D43E}{\relax\ifmmode K \else $K$\fi}
\DeclareUnicodeCharacter{1D43F}{\relax\ifmmode L \else $L$\fi}
\DeclareUnicodeCharacter{1D440}{\relax\ifmmode M \else $M$\fi}
\DeclareUnicodeCharacter{1D441}{\relax\ifmmode N \else $N$\fi}
\DeclareUnicodeCharacter{1D442}{\relax\ifmmode O \else $O$\fi}
\DeclareUnicodeCharacter{1D443}{\relax\ifmmode P \else $P$\fi}
\DeclareUnicodeCharacter{1D444}{\relax\ifmmode Q \else $Q$\fi}
\DeclareUnicodeCharacter{1D445}{\relax\ifmmode R \else $R$\fi}
\DeclareUnicodeCharacter{1D446}{\relax\ifmmode S \else $S$\fi}
\DeclareUnicodeCharacter{1D447}{\relax\ifmmode T \else $T$\fi}
\DeclareUnicodeCharacter{1D448}{\relax\ifmmode U \else $U$\fi}
\DeclareUnicodeCharacter{1D449}{\relax\ifmmode V \else $V$\fi}
\DeclareUnicodeCharacter{1D44A}{\relax\ifmmode W \else $W$\fi}
\DeclareUnicodeCharacter{1D44B}{\relax\ifmmode X \else $X$\fi}
\DeclareUnicodeCharacter{1D44C}{\relax\ifmmode Y \else $Y$\fi}
\DeclareUnicodeCharacter{1D44D}{\relax\ifmmode Z \else $Z$\fi}
\DeclareUnicodeCharacter{1D44E}{\relax\ifmmode a \else $a$\fi}
\DeclareUnicodeCharacter{1D44F}{\relax\ifmmode b \else $b$\fi}
\DeclareUnicodeCharacter{1D450}{\relax\ifmmode c \else $c$\fi}
\DeclareUnicodeCharacter{1D451}{\relax\ifmmode d \else $d$\fi}
\DeclareUnicodeCharacter{1D452}{\relax\ifmmode e \else $e$\fi}
\DeclareUnicodeCharacter{1D453}{\relax\ifmmode f \else $f$\fi}
\DeclareUnicodeCharacter{1D454}{\relax\ifmmode g \else $g$\fi}
\DeclareUnicodeCharacter{1D456}{\relax\ifmmode i \else $i$\fi}
\DeclareUnicodeCharacter{1D457}{\relax\ifmmode j \else $j$\fi}
\DeclareUnicodeCharacter{1D458}{\relax\ifmmode k \else $k$\fi}
\DeclareUnicodeCharacter{1D459}{\relax\ifmmode l \else $l$\fi}
\DeclareUnicodeCharacter{1D45A}{\relax\ifmmode m \else $m$\fi}
\DeclareUnicodeCharacter{1D45B}{\relax\ifmmode n \else $n$\fi}
\DeclareUnicodeCharacter{1D45C}{\relax\ifmmode o \else $o$\fi}
\DeclareUnicodeCharacter{1D45D}{\relax\ifmmode p \else $p$\fi}
\DeclareUnicodeCharacter{1D45E}{\relax\ifmmode q \else $q$\fi}
\DeclareUnicodeCharacter{1D45F}{\relax\ifmmode r \else $r$\fi}
\DeclareUnicodeCharacter{1D460}{\relax\ifmmode s \else $s$\fi}
\DeclareUnicodeCharacter{1D461}{\relax\ifmmode t \else $t$\fi}
\DeclareUnicodeCharacter{1D462}{\relax\ifmmode u \else $u$\fi}
\DeclareUnicodeCharacter{1D463}{\relax\ifmmode v \else $v$\fi}
\DeclareUnicodeCharacter{1D464}{\relax\ifmmode w \else $w$\fi}
\DeclareUnicodeCharacter{1D465}{\relax\ifmmode x \else $x$\fi}
\DeclareUnicodeCharacter{1D466}{\relax\ifmmode y \else $y$\fi}
\DeclareUnicodeCharacter{1D467}{\relax\ifmmode z \else $z$\fi}

\DeclareUnicodeCharacter{1E67}{\.{\v s}}
\DeclareUnicodeCharacter{1E11}{\relax\ifmmode \c{d} \else $\c{d}$\fi}
\DeclareUnicodeCharacter{1ECB}{\relax\ifmmode \d{i} \else $\d{i}$\fi}
\DeclareUnicodeCharacter{1D8D}{\relax\ifmmode \textlhookx \else $\textlhookx$\fi}
\DeclareUnicodeCharacter{104}{\relax\ifmmode \k{A} \else $\k{A}$\fi}
\DeclareUnicodeCharacter{211E}{\relax\ifmmode \textrecipe \else $\textrecipe$\fi}
\DeclareUnicodeCharacter{29D}{\relax\ifmmode \textctj \else $\textctj$\fi}

\DeclareUnicodeCharacter{1E2E}{\'{\"I}}
\DeclareUnicodeCharacter{23F}{\textrts}

\DeclareUnicodeCharacter{2C73}{\varw}

\DeclareUnicodeCharacter{2127}{\mho}

\DeclareUnicodeCharacter{28C}{\textturnv}
\DeclareUnicodeCharacter{252}{\textturnscripta}
\DeclareUnicodeCharacter{259}{\schwa}
\DeclareUnicodeCharacter{25B}{\m{e}}
\DeclareUnicodeCharacter{266}{\m{h}}
\DeclareUnicodeCharacter{127}{\B{h}}
\DeclareUnicodeCharacter{27E}{\textfishhookr}
\DeclareUnicodeCharacter{281}{\textinvscr}

\firstpage{1}
\articlenumber{838}
\pubvolume{16}
\issuenum{6}
\pubyear{2026}
\copyrightyear{2026}
\externaleditor{Diego~Vergara} 
\datereceived{13 April 2026}
\daterevised{23 May 2026}
\dateaccepted{24 May 2026}
\datepublished{27 May 2026}

\usepackage[OT1,OT2,T2A,T2B,T2C,T3,T5,T1]{fontenc}
\usepackage[russian,english]{babel}

\Title{Multi-Level Barriers to Generative AI Adoption Across Disciplines and Professional Roles in Higher Education}

\Author{Jianhua~Yang~\textsuperscript{1}\textsuperscript{,}*\href{https://orcid.org/0000-0002-6873-7653}{\orcidicon}, Kerem~Öge~\textsuperscript{2}\href{https://orcid.org/0000-0002-3603-1058}{\orcidicon}, Adrian~von Mühlenen~\textsuperscript{3}\href{https://orcid.org/0000-0001-8338-8554}{\orcidicon}, Abdullah Bilal~Akbulut~\textsuperscript{4}\href{https://orcid.org/0009-0008-4955-7770}{\orcidicon}, Tanya Suzanne~Carey~\textsuperscript{1} and~Chidi~Okorro~\textsuperscript{1}\href{https://orcid.org/0009-0001-0126-5284}{\orcidicon}}

\AuthorNames{Jianhua Yang, Kerem Öge, Adrian von Mühlenen, Abdullah Bilal Akbulut, Tanya Suzanne Carey and Chidi Okorro}

\address{\textsuperscript{1} \quad Warwick Manufacturing Group, The University of Warwick, Coventry CV4 7AL, UK; t.carey@warwick.ac.uk~(T.S.C.); chidi.okorro@warwick.ac.uk~(C.O.)

\textsuperscript{2} \quad Department of Politics and International Studies, The University of Warwick, Coventry CV4 7AL, UK; kerem.oge@warwick.ac.uk

\textsuperscript{3} \quad Department of Psychology, The University of Warwick, Coventry CV4 7AL, UK; a.vonmuhlenen@warwick.ac.uk

\textsuperscript{4} \quad Birmingham Business School, The University of Birmingham, Birmingham B15 2TT, UK; axa2301@student.bham.ac.uk}

\corres{Correspondence: jianhua.yang@warwick.ac.uk}

\abstract{Generative artificial intelligence (GenAI) is rapidly reshaping higher education, yet barriers to its adoption across different disciplines and institutional roles remain underexplored. The existing literature frequently attributes adoption barriers to individual-level factors such as perceived usefulness and ease of use. This study instead investigates how such barriers are associated with structural conditions. Drawing on a multi-method survey analysis of 272 academic and professional service (PS) staff at Russell Group university, we examine how disciplinary contexts and institutional roles influence perceived barriers. By integrating multinomial logistic regression (MLR), structural equation modelling (SEM), and semantic clustering of open-ended responses, we move beyond descriptive accounts to develop a multi-level account of GenAI adoption. Our findings reveal patterned differences: non-STEM academics primarily report ethical and cultural barriers related to academic integrity, whereas STEM and PS staff disproportionately emphasize institutional, governance, and infrastructure constraints. We conclude that GenAI adoption barriers are deeply embedded in organizational ecosystems and epistemic norms, while also reflecting individual experiences and other unmeasured factors, suggesting that universities must move beyond generalized training to develop role-specific governance and \mbox{support frameworks}.}


\makeatletter
\DeclareRobustCommand*\textsubscript[1]{%
  \@textsubscript{\selectfont#1}}
\def\@textsubscript#1{%
  {\m@th\ensuremath{_{\mbox{\fontsize\sf@size\z@#1}}}}}
\makeatother

\usepackage{sansmath}

\begin{document}
\section{Introduction \label{sect:sec1-education-4286608}}

Generative artificial intelligence (GenAI) is rapidly reshaping higher education (HE), offering transformative opportunities for teaching and learning while posing significant concerns (\citealp{B20-education-4286608}). Large language models (LLMs) are now used in teaching preparation, assessment design, student support, administration, and institutional communication. They promise efficiency and pedagogical innovation (\citealp{B22-education-4286608}), yet they also raise concerns about academic integrity, bias, privacy, workload, and the erosion of core academic skills (\citealp{B1-education-4286608}; \citealp{B51-education-4286608}). Adoption is therefore uneven and often contested across universities.

Many studies explain technology adoption primarily through individual-level factors such as perceived usefulness, ease of use, literacy, or attitudes (\citealp{B29-education-4286608}; \citealp{B52-education-4286608}; \citealp{B59-education-4286608}). While valuable, these approaches understate the institutional complexity of HE. Universities are structured by disciplinary norms, professional identities, governance arrangements, and regulatory constraints. Decisions about GenAI use are embedded within institutional roles and subject cultures.

This study asks how perceived barriers to GenAI adoption in HE are associated with individual-level issues, institutional roles, disciplines, norms, and ethical expectations. This distinction has direct policy implications. If barriers are individual, training interventions may suffice. If they are structural, governance design and role-specific support \mbox{become central.}

To address this question, we draw on a 2025 survey of staff at Russell Group university on GenAI use and barriers, completed by 272 respondents. The survey covers both academic and professional service (PS) roles and captures literacy, attitudes, job threat perceptions, institutional guidance and support, ethical concerns, and perceived barriers across teaching and workplace contexts. Methodologically, we adopt a multi-method approach to survey analysis. We combine descriptive statistics, multinomial logistic regression, structural equation modelling (SEM), and the clustering of free-text responses to analyze both systematic differences and underlying mechanisms. This design allows us to connect reported barriers to disciplines and professional roles.

Our findings show clear and systematic differences in reported barriers across disciplines and roles. STEM (science, technology, engineering, and mathematics) staff are more likely to report institutional and individual barriers, whereas non-STEM staff report higher ethical and cultural barriers in absolute terms. PS staff differ significantly from academics, particularly in relation to institutional constraints they face. We also find that individual factors such as literacy, attitudes, and job threat matter, but they operate within structured disciplinary and organizational contexts. These results suggest that GenAI adoption barriers reflect academic disciplinary cultures, professional roles, and ethical expectations alongside individual skill or perception.

Our contribution is threefold. First, we provide one of the few systematic cross-disciplinary analyses of GenAI barriers, directly comparing STEM and non-STEM contexts within the same institution. Second, and more distinctively, we extend the analysis beyond teaching academics to include PS staff. This group plays a critical role in university operations yet is largely absent from the GenAI adoption literature. To our knowledge, no prior study offers a large-scale comparative analysis of both academic and PS roles within a single-institutional setting. Third, by integrating quantitative modelling with qualitative clustering, we move beyond descriptive accounts of barriers and develop a multi-level, empirically grounded account of GenAI adoption. The remainder of the paper proceeds as follows: the next section reviews relevant theoretical and empirical literature, followed by a description of the research design and methods; we then present the results and a discussion of our findings.

\section{Barriers to GenAI Adoption in HE \label{sect:sec2-education-4286608}}

GenAI has generated both optimism and unease in HE. Advocates argue that it may significantly reshape teaching and learning practices, enhancing feedback, personalization, and efficiency. Critics, however, emphasize risks to academic integrity, authorship, and professional standards (\citealp{B14-education-4286608}). Some scholars contend that many of these risks stem not from the technology itself but from assessment systems designed for a pre-GenAI environment and therefore call for redesign rather than prohibition (\citealp{B8-education-4286608}; \citealp{B26-education-4286608}). Others warn that habitual reliance on GenAI may weaken the development of critical thinking and writing skills (\citealp{B40-education-4286608}). While much of this debate is framed normatively, empirical research increasingly points to a heterogeneous set of barriers shaping adoption in practice (\citealp{B13-education-4286608}).

To organize this literature, we distinguish between individual, ethical, socio-cultural, and institutional barriers. This distinction refers primarily to the level at which barriers are experienced and reported, while recognizing that their underlying causes may overlap across categories. Individual barriers refer to constraints articulated at the level of the user, including limited knowledge, confidence, and time. As mentioned above, it is important to note that these are not necessarily personal deficits, since time constraints or low confidence often reflect workload pressures or insufficient support. Ethical barriers refer to concerns about whether GenAI use is appropriate, legitimate, or responsible. Socio-cultural barriers refer to shared norms and expectations embedded in disciplinary communities. Institutional barriers refer to constraints produced by organizational systems, including policy ambiguity, infrastructure, and access to approved tools. This distinction extends beyond standard technology adoption models such as the Technology Acceptance Model (TAM) and Unified Theory of Acceptance and Use of Technology (UTAUT), which explain adoption through perceived usefulness, ease of use, intention, literacy, and attitudes, but pay less attention to how these perceptions are shaped by institutional and \mbox{disciplinary contexts.}

We begin with individual barriers. Classic individual-level technology adoption frameworks such as the TAM emphasize perceived usefulness and ease of use as key determinants of uptake (\citealp{B23-education-4286608}; \citealp{B54-education-4286608}; \citealp{B59-education-4286608}). These constructs remain highly relevant in GenAI contexts, particularly in explaining variations in literacy, confidence, and perceptions of effectiveness (\citealp{B2-education-4286608}). Empirical studies identify a range of individual-level constraints, including limited AI literacy among staff and students (\citealp{B41-education-4286608}), insufficient training and low self-efficacy, time pressures that restrict experimentation, and concerns about inaccuracy or misinformation (\citealp{B48-education-4286608}). Yet emerging scholarship cautions against reducing barriers to individual cognition alone. Models centred narrowly on attitudes risk overlooking the organizational, cultural, and societal conditions that shape how such perceptions are formed and acted upon (\citealp{B11-education-4286608}; \citealp{B27-education-4286608}; \citealp{B49-education-4286608}).

Second, a substantial body of research demonstrates that resistance to AI is frequently grounded in ethical and socio-cultural concerns rather than purely technical limitations (\citealp{B6-education-4286608}; \citealp{B19-education-4286608}; \citealp{B25-education-4286608}; \citealp{B35-education-4286608}; \citealp{B38-education-4286608}), which are dimensions that are only partially captured within UTAUT style models (\citealp{B46-education-4286608}). Documented concerns include privacy and data security (\citealp{B16-education-4286608}; \citealp{B28-education-4286608}), academic integrity and plagiarism (\citealp{B51-education-4286608}), intellectual property rights (\citealp{B10-education-4286608}), pedagogical orientation (\citealp{B15-education-4286608}), skill erosion and over-reliance (\citealp{B17-education-4286608}), and threats to intellectual authenticity and creativity (\citealp{B48-education-4286608}). Scholars also highlight anxieties about dehumanization and the erosion of professional judgement, distrust in AI outputs, ethical ambivalence regarding training data and monetization (\citealp{B57-education-4286608}), inequitable access (\citealp{B34-education-4286608}; \citealp{B53-education-4286608}), and cross-cultural variation in acceptance (\citealp{B56-education-4286608}; \citealp{B60-education-4286608}). Taken together, these concerns indicate that GenAI unsettles normative understandings of academic labour and professional responsibility. Adoption decisions are therefore shaped not only by instrumental evaluations of utility, but also by ethical expectations, socio-cultural contexts, disciplinary norms, and \mbox{professional identities.}

Beyond these barriers, institutional structures and norms play a decisive role in shaping adoption conditions. A growing body of research identifies structural barriers such as the absence of clear institutional guidance (\citealp{B7-education-4286608}; \citealp{B44-education-4286608}; \citealp{B45-education-4286608}), limited development of inclusive governance frameworks, and cost and infrastructure constraints, including bandwidth limitations and digital divides. Studies also point to gendered differences in access (\citealp{B5-education-4286608}; \citealp{B50-education-4286608}), compliance pressures linked to data protection regimes such as GDPR (General Data Protection Regulation) and the EU AI Act, and the need for formal oversight and ethics governance mechanisms (\citealp{B33-education-4286608}). Even where high-level policy encouragement exists (\citealp{B18-education-4286608}), translating broad guidance into local, department-level practice remains challenging (\citealp{B39-education-4286608}). Taken together, these findings suggest that GenAI adoption is embedded within governance environments and institutional capacities. Perceived barriers may therefore reflect structural constraints and organizational design rather than simple individual reluctance.

Recent scholarship argues that adoption in complex educational settings should be analyzed through multi-level frameworks that integrate individual, ethical, and institutional dimensions rather than isolating beliefs or attitudes (\citealp{B11-education-4286608}; \citealp{B27-education-4286608}; \citealp{B49-education-4286608}). Yet, two important gaps remain. First, cross-disciplinary comparisons are still relatively rare (\citealp{B62-education-4286608}). Second, the PS workforce is largely absent from empirical analyses, despite its central role in governance, compliance, infrastructure, and operational workflows (\citealp{B37-education-4286608}). This omission constrains our ability to distinguish clearly between barriers rooted in individual cognition and those produced by structural or organizational conditions. By systematically comparing STEM and non-STEM contexts and incorporating PS staff alongside academics, the present study addresses these gaps and advances the literature from a descriptive catalogue of concerns towards a structured, multi-level explanation of GenAI adoption in HE.

\section{Hypotheses \label{sect:sec3-education-4286608}}

Building on recent scholarship, we contend that GenAI adoption in HE cannot be adequately explained by individual characteristics alone, such as perceived usefulness, ease of use, or technical literacy. Although these factors remain important, accumulating evidence shows that adoption decisions are embedded within wider ethical considerations (\citealp{B12-education-4286608}), disciplinary traditions, professional norms, and institutions. These contextual forces shape not only access and capability, but also how staff interpret the risks, responsibilities, and opportunities associated with GenAI. Understanding adoption therefore requires moving beyond a purely cognitive account towards a framework that situates individual perceptions within organizational and disciplinary environments.

Disciplinary context is particularly salient in shaping GenAI adoption. Academic fields are structured around distinct epistemic norms, assessment formats, and standards of evidence (\citealp{B55-education-4286608}). In writing-intensive and interpretive disciplines, concerns about authorship, originality, and academic integrity may be especially pronounced. By contrast, in technically oriented disciplines, issues such as model accuracy, reproducibility, and integration into established workflows may take precedence. Prior research points to cross-disciplinary variation in perceptions of fairness, accountability, and pedagogical alignment (\citealp{B42-education-4286608}; \citealp{B61-education-4286608}). Although some cross-disciplinary evidence exists for students, including reported differences across STEM and non-STEM cohorts and gendered patterns of attitudes (\citealp{B9-education-4286608}; \citealp{B36-education-4286608}), systematic empirical comparisons among staff remain limited. This gap restricts our ability to assess whether disciplinary norms shape barrier perceptions in comparable ways across professional roles within universities.

Institutional role constitutes a second structural dimension shaping GenAI adoption. Academic staff are directly responsible for teaching, assessment, and disciplinary knowledge production, whereas PS staff operate within governance, compliance, and operational infrastructures. Their exposure to GenAI therefore differs not only in function but also in perceived risk and accountability. PS staff are more likely to encounter barriers linked to data governance, regulatory compliance, infrastructure, and workflow automation. Academics, by contrast, may foreground concerns about pedagogical integrity, authorship, student capability development, and academic standards.

Empirical evidence supports this role-based differentiation and demonstrates that institutional position shapes how GenAI is perceived and problematized in HE. Educators tend to frame GenAI primarily in pedagogical and epistemic terms, focusing on assessment integrity, student over-reliance, disciplinary standards, and workload implications. Administrators, in contrast, emphasize governance coherence, data privacy, compliance obligations, and institutional risk management. Other studies show that educators, students, and administrators face role-specific barriers, and that professional and administrative staff should be analyzed as a distinct group (\citealp{B31-education-4286608}; \citealp{B32-education-4286608}). There is also evidence that non-academic staff are frequently excluded from policy formation processes (\citealp{B21-education-4286608}). Taken together, this literature suggests that perceived barriers to GenAI adoption are likely to vary systematically between academic and administrative roles, reflecting differences in responsibility and institutional positioning.

The existing literature indicates that institutional position and disciplinary context systematically shape how staff perceive barriers to GenAI adoption. Rather than treating adoption as the outcome of individual attitudes or technical capability alone, recent work suggests that staff interpret risks and opportunities through their location within organizational hierarchies and epistemic communities. These structural positions influence which constraints become salient and which forms of GenAI use appear legitimate, risky, or valuable. To examine this proposition empirically, we formulate the following hypothesis:

\vspace{12pt}
\noindent \textbf{\boldmath{H1.}}  \emph{Perceived barriers to GenAI adoption in HE vary systematically across institutional roles and disciplinary contexts, beyond individual-level attitudes and technical capability.}

\vspace{12pt}
We specify this hypothesis in two dimensions:

\vspace{12pt}
\noindent \textbf{\boldmath{H1a.}}  \emph{Staff working in STEM and non-STEM disciplines will report systematically different \mbox{GenAI barriers}.}

\vspace{12pt}
\noindent \textbf{\boldmath{H1b.}}  \emph{Academic staff and PS staff will exhibit systematically different GenAI barriers.}


\section{Methodology \label{sect:sec4-education-4286608}}

We examine barriers to GenAI adoption through an in-depth single-institution case study at Russell Group university. This allows us to analyze variation across disciplinary contexts and institutional roles within a shared governance and policy environment, thereby isolating patterned differences in perceived barriers. We collected data through a cross-sectional survey combining structured quantitative measures with open-ended qualitative responses (\citealp{B24-education-4286608}).

Similarly to recent studies examining university staff’s perceptions of GenAI use (\citealp{B58-education-4286608}), the current study employed a voluntary, non-probability sampling approach, rather than attempting to obtain a statistically representative sample of all staff at the institution. Given the rapidly evolving nature of GenAI adoption in HE, early-stage studies like the present prioritize timely and contextual insight over probabilistic representativeness. In addition, disciplinary grouping was operationalised using a STEM vs. non-STEM distinction. It is acknowledged that this binary classification is imperfect, as joint degrees, interdisciplinary programmes, and computationally oriented areas within the social sciences increasingly blur the boundaries between traditional disciplines. However, within many HE institutions, including the current Russell Group university, teaching, administration, and organizational structures remain substantially organized at faculty or school level along broad STEM and non-STEM divisions. Furthermore, the STEM vs. non-STEM distinction remains a common analytical grouping in HE and educational technology research, particularly in exploratory studies where sample size constraints limit more fine-grained disciplinary comparisons (\citealp{B3-education-4286608}).

We developed the survey instrument following the above systematic review of the GenAI literature. The survey was organized around GenAI usage in teaching and professional work, broader perceptions of impact, institutional policy and support, and demographic characteristics. Core barriers in the literature were translated into measurable items capturing institutional role, disciplinary location, GenAI literacy, attitudes towards GenAI, perceived job threat, perceived institutional guidance and support, current and intended GenAI use, and perceived barriers to adoption, where most perceptual variables were measured using five-point Likert scales (\citealp{B24-education-4286608}) as shown in \tabref{tabref:education-4286608-t001}.    
    \begin{table}[H]
    \footnotesize
    
    \caption{Questions in the staff survey, where dependent variables have options falling into categories of Ind = Individual, Ins = Institutional, E = Ethical, and C = Cultural.}
    \label{tabref:education-4286608-t001}

\begin{adjustwidth}{-\extralength}{0cm}
\setlength{\cellWidtha}{\fulllength/4-2\tabcolsep-0.9in}
\setlength{\cellWidthb}{\fulllength/4-2\tabcolsep+0.5in}
\setlength{\cellWidthc}{\fulllength/4-2\tabcolsep+1in}
\setlength{\cellWidthd}{\fulllength/4-2\tabcolsep-0.6in}
\scalebox{1}[1]{\begin{tabularx}{\fulllength}{>{\raggedright\arraybackslash}m{\cellWidtha}>{\raggedright\arraybackslash}m{\cellWidthb}>{\raggedright\arraybackslash}m{\cellWidthc}>{\raggedright\arraybackslash}m{\cellWidthd}}
\toprule

\multicolumn{1}{>{\centering\arraybackslash}m{\cellWidtha}}{} & \multicolumn{1}{>{\centering\arraybackslash}m{\cellWidthb}}{\textbf{Question}} & \multicolumn{1}{>{\centering\arraybackslash}m{\cellWidthc}}{\textbf{Options}} & \multicolumn{1}{>{\centering\arraybackslash}m{\cellWidthd}}{\textbf{Note}}\\
\cmidrule{1-4}

\multirow{2}{*}{\parbox{\cellWidtha}{\raggedright Independent variables}} & In which department or unit do you work at the University? & Combo box (dropdown selection plus one line text~box) & STEM or non-STEM\\

 & Which of the following best describes your current role? & Academic; Professional Services & Job Role\\
\cmidrule{1-4}
Dependent variables & What are the main barriers to using GenAI in your teaching/work? & Lack of institutional guidance or policies (Ins); \linebreak Concerns about academic integrity and student misuse~(E); \linebreak Limited personal knowledge or training on AI tools~(Ind); \linebreak Lack of time to explore and implement AI solutions~(Ind); \linebreak Ethical concerns about bias, privacy, or surveillance (E); \linebreak Uncertainty about AI’s effectiveness in improving learning outcomes (Ind); \linebreak Technical difficulties or lack of institutional support~(Ins); \linebreak Concerns about AI replacing human elements in teaching (E); \linebreak Resistance from colleagues or institutional culture (C); \linebreak Lack of funding or access to appropriate AI tools (Ins); \linebreak Preference for traditional teaching methods (C) & Select three most important.\\
\cmidrule{1-4}
\multirow{5}{*}{\parbox{\cellWidtha}{\raggedright Mediator}} & How would you describe your current level of GenAI literacy? & Likert scale 1--5, no understanding to expert & Literacy\\

 & What is your overall view of GenAI in~education? & Likert scale 1--5, very negative to very positive & Attitude\\

 & I am concerned that GenAI will become a threat to my job in the next 5 years. & Likert scale 1--5, strongly disagree to strongly agree & Job Threat\\

 & My institution/academic department has provided clear guidance on the acceptable and unacceptable uses of GenAI in teaching/work. & Likert scale 1--5, strongly disagree to strongly agree & Guidance\\

 & My institution/academic department has provided sufficient resources to develop staff GenAI literacy. & Likert scale 1--5, strongly disagree to strongly agree & Support\\
\cmidrule{1-4}
Free-entry texts & What are the main barriers to using GenAI in your teaching/work? & Free-text inputs &  \\

\bottomrule
\end{tabularx}}
\end{adjustwidth}

    \end{table}

Multinomial logistic regression (MLR) was conducted using the Python 3.12 statistical modelling library statsmodels (\citealp{B47-education-4286608}). To capture barrier structures, respondents selected the three most salient barriers from a predefined list spanning individual, institutional, ethical, and cultural dimensions. For the MLR analysis, these selections were analyzed at the barrier-category occurrence level: each selected barrier category contributed one occurrence to the nominal outcome. This is equivalent to analyzing a long-format dataset in which repeated category selections are represented as multiple observations. This operationalisation preserves information from all selected barrier categories and avoids reducing each respondent to a single modal or dominant barrier type. However, because up to three barrier-category occurrences could come from the same respondent, observations are not fully independent; the results should therefore be interpreted as category-level selection patterns rather than respondent-level classifications.

\begin{figure}[H]
      \includegraphics[scale=1]{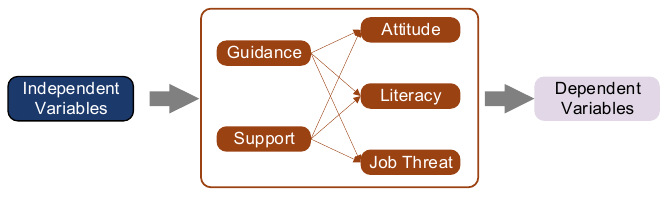}
\caption{\textls[-25]{Conceptual model showing interacting mediators linking independent and \mbox{dependent variables}.}}
\label{fig:education-4286608-f001}
\end{figure}

We developed a conceptual model in which independent variables influence a set of mediators, which in turn shape the dependent variables, namely different levels of GenAI adoption barriers, as shown in \fig{fig:education-4286608-f001}. Within the mediating layer, we propose that institutional factors shape individual perspectives, including attitudes, AI literacy, and perceived job threat, which in turn influence the level and intensity of perceived barriers. The model is theoretically informed by technology adoption and organizational-readiness literature (\citealp{B30-education-4286608}; \citealp{B54-education-4286608}); however, it is used here as an exploratory path model rather than as a confirmed causal ordering. Guidance and support are positioned as institutional conditions that may shape literacy, attitudes, and perceived job threat, while the possibility that they may also act directly on perceived barriers is acknowledged as both a theoretical and empirical consideration. Such a model also enables investigation through quantitative approaches such as SEM, which was conducted using SPSS Amos 31.    

The survey included free-text entries to capture barriers not covered by the predefined items. We analyzed these open-ended responses using embedding-based semantic clustering. Specifically, responses were first embedded with OpenAI’s text-embedding-3-large model in a 3072-dimensional semantic vector space and reduced with UMAP (uniform manifold approximation and projection) (\citealp{B4-education-4286608}) before clustering to preserve local semantic neighbourhoods while reducing sparsity (cosine similarity, local neighbourhood size = 15, output dimensions = 20). HDBSCAN (hierarchical density-based spatial clustering of applications with noise) (\citealp{B43-education-4286608}) was then used for clustering because it can recover uneven thematic groups while leaving weakly connected responses unclustered rather than forcing all texts into a fixed number of categories. The clustering settings were deliberately permissive to retain smaller but interpretable themes without excessive fragmentation: Euclidean distance on the reduced vectors and minimum thematic cluster sizes = 3, allowing single-point clusters. In choosing these parameters, attention was paid to maintaining a balance between reducing unclustered responses while also penalizing highly uneven cluster sizes. Cluster labels were then generated using OpenAI’s gpt-5-mini model from up to 10 exemplar responses nearest to each cluster centroid for the 10 largest clusters. The generated labels were subsequently reviewed manually against the exemplar texts and representative phrases before interpretation. This unsupervised approach allows thematic structures to emerge from the data rather than imposing predefined coding categories.

By integrating regression analysis, structural modelling, and semantic clustering within a shared institutional context, we move from identifying patterned differences in barrier profiles to examining the mechanisms that shape them.

\section{Results \label{sect:sec5-education-4286608}}

We implemented the survey in Qualtrics using role-based branching to ensure relevance for both academic and PS staff. Following ethical approval, we administered the survey between mid-June and late July 2025, relying on non-probability recruitment through institutional communications and departmental dissemination. The final dataset comprises 272 valid responses from 72 STEM and 200 non-STEM staff members, of whom 148 were academics and 124 were PS staff.

\subsection{Barrier Distribution by Job Role and Discipline \label{sect:sec5dot1-education-4286608}}

\begin{figure}[H]
      \includegraphics[scale=1]{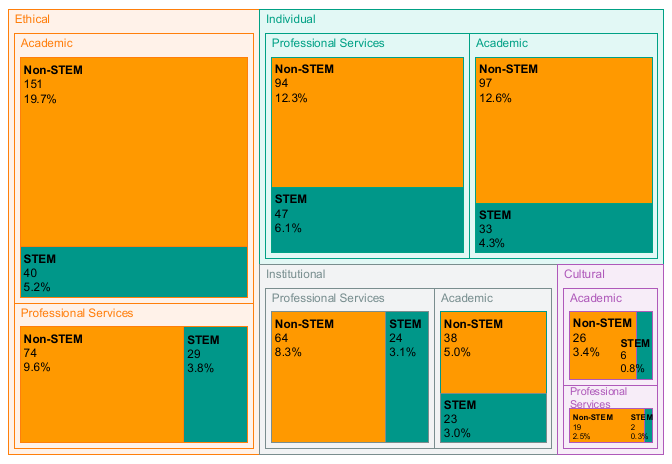}
\caption{Barrier percentages differentiating job roles and disciplines shown in a treemap.}
\label{fig:education-4286608-f002}
\end{figure}

\fig{fig:education-4286608-f002} presents the distribution of barrier categories across disciplinary context and institutional role. The descriptive patterns suggest patterned variation across the four role-by-discipline groups.    

First, non-STEM teaching staff report ethical barriers more often than other groups. Non-STEM academics account for the largest share of concerns related to academic integrity, over-reliance, and the erosion of critical thinking. By contrast, STEM respondents report ethical barriers less frequently. This pattern aligns with disciplinary differences in epistemic norms and assessment formats, where writing-intensive and interpretive disciplines face more immediate challenges to authorship and originality.

Second, compared to academic staff, institutional barriers are more prevalent among PS staff, particularly in non-STEM administrative units. These include concerns related to licencing restrictions, policy ambiguity, governance processes, and infrastructure constraints. This pattern suggests that PS staff experience GenAI more through organizational systems and compliance structures rather than through classroom pedagogy. Third, individual barriers such as lack of training, limited time, and uncertainty about effectiveness are more evenly distributed across groups but remain discipline-sensitive. While present in both STEM and non-STEM contexts, they do not dominate any single structural position.

Finally, results show that barrier profiles differ markedly by institutional role. Among academic staff, ethical barriers are the most prominent category, especially within non-STEM disciplines, where concerns about academic integrity, student over-reliance, and the erosion of critical thinking dominate. In contrast, PS staff report comparatively fewer ethical concerns and instead display a stronger concentration of individual and institutional barriers, including limitations related to training, time, licencing, infrastructure, and governance. While disciplinary variation persists within both groups, the overall pattern suggests a role-based shift in how GenAI is problematized: academics frame GenAI primarily as an epistemic and pedagogical risk, whereas PS staff experience it more as a capability and organizational implementation challenge.

These descriptive differences provide initial descriptive evidence for H1a and H1b: perceived barriers vary systematically across both disciplinary contexts and \mbox{institutional roles.}

\subsection{Multinomial Regression of Barrier-Category Membership \label{sect:sec5dot2-education-4286608}}

To assess whether the descriptive differences reflect systematic structural effects, we used MLR to model the category distribution of selected barriers. We used cultural barriers as the reference category, allowing ethical, individual, and institutional barriers to be interpreted relative to cultural barriers. Because respondents could select more than one barrier, odds ratios refer to selected barrier categories rather than mutually exclusive respondent profiles. Cultural barriers, as the reference category, are therefore absent as a separate coefficient row in the comparison tables.

The model including STEM status was statistically significant, as seen in \tabref{tabref:education-4286608-t002}, indicating that disciplinary affiliation is associated with the distribution of selected barrier categories. However, pseudo-\emph{R}\textsuperscript{2} values were small (Cox \& Snell = 0.011; \mbox{Nagelkerke = 0.012;} McFadden = 0.004), suggesting that discipline explains only a limited proportion of \mbox{overall variance.}    
    \begin{table}[H]
    
    \caption{MLR model fit between STEM/non-STEM and barriers, where STEM = 1 is the predictor reference group.}
    \label{tabref:education-4286608-t002}

\setlength{\cellWidtha}{\textwidth/5-2\tabcolsep-0in}
\setlength{\cellWidthb}{\textwidth/5-2\tabcolsep+0.6in}
\setlength{\cellWidthc}{\textwidth/5-2\tabcolsep-0.2in}
\setlength{\cellWidthd}{\textwidth/5-2\tabcolsep-0.2in}
\setlength{\cellWidthe}{\textwidth/5-2\tabcolsep-0.2in}
\scalebox{1}[1]{\begin{tabularx}{\textwidth}{>{\centering\arraybackslash}m{\cellWidtha}>{\centering\arraybackslash}m{\cellWidthb}>{\centering\arraybackslash}m{\cellWidthc}>{\centering\arraybackslash}m{\cellWidthd}>{\centering\arraybackslash}m{\cellWidthe}}
\toprule

\textbf{Model} & \textbf{$\boldmath{-}$2 Log Likelihood} & \textbf{LR $\bm{\upchi}$\textsuperscript{2}} & \textbf{df} & \textbf{\emph{p}}\\
\cmidrule{1-5}

Intercept-only & 41.569 &   &   &  \\

Final & 33.056 & 8.512 & 3 & 0.037\\

\bottomrule
\end{tabularx}}

    \end{table}

At the category level, compared with cultural barriers, non-STEM respondents had significantly lower odds of being classified into the individual and institutional categories (\tabref{tabref:education-4286608-t003}). Substantively, this indicates that STEM respondents are more likely than non-STEM respondents to interpret barriers in terms of capability and infrastructure constraints, that is, individual and institutional barriers, rather than as forms of cultural resistance. The contrast for ethical barriers was not statistically significant. Disciplinary context therefore shapes how selected barriers are distributed, but its effects are selective rather than uniform across all categories.    
    \begin{table}[H]
    
    \caption{STEM group effects by selected barrier category (reference outcome = cultural; reference predictor group = STEM).}
    \label{tabref:education-4286608-t003}

\begin{adjustwidth}{-\extralength}{0cm}
\setlength{\cellWidtha}{\fulllength/5-2\tabcolsep-0in}
\setlength{\cellWidthb}{\fulllength/5-2\tabcolsep-0.3in}
\setlength{\cellWidthc}{\fulllength/5-2\tabcolsep-0.3in}
\setlength{\cellWidthd}{\fulllength/5-2\tabcolsep-0.3in}
\setlength{\cellWidthe}{\fulllength/5-2\tabcolsep+0.9in}
\scalebox{1}[1]{\begin{tabularx}{\fulllength}{>{\centering\arraybackslash}m{\cellWidtha}>{\centering\arraybackslash}m{\cellWidthb}>{\centering\arraybackslash}m{\cellWidthc}>{\centering\arraybackslash}m{\cellWidthd}>{\centering\arraybackslash}m{\cellWidthe}}
\toprule

\textbf{Outcome Category} & \textbf{B} & \textbf{OR = Exp(B)} & \textbf{\emph{p}} & \textbf{Notes}\\
\cmidrule{1-5}

Ethical & $-$0.545 & 0.58 & 0.181 & Not significant\\

Individual & $-$0.857 & 0.424 & 0.035 & non-STEM lower odds than STEM\\

Institutional & $-$0.952 & 0.386 & 0.024 & non-STEM lower odds than STEM\\

\bottomrule
\end{tabularx}}
\end{adjustwidth}

    \end{table}

A second multinomial model examined the effect of job role, with academic staff as the reference group (academic = 1) and PS staff as the comparison group. The model was highly significant (as seen in \tabref{tabref:education-4286608-t004}), with somewhat larger pseudo-\emph{R}\textsuperscript{2} values (\mbox{Cox \& Snell = 0.038}; Nagelkerke = 0.041; McFadden = 0.016), indicating that institutional role is more strongly associated with barrier-category distribution than discipline, although it still has modest explanatory power.    
    \begin{table}[H]
    
    \caption{MLR model fit between academic/non-academic and barriers, where academic = 1 is the predictor reference group.}
    \label{tabref:education-4286608-t004}

\setlength{\cellWidtha}{\textwidth/5-2\tabcolsep-0in}
\setlength{\cellWidthb}{\textwidth/5-2\tabcolsep+0.6in}
\setlength{\cellWidthc}{\textwidth/5-2\tabcolsep-0.2in}
\setlength{\cellWidthd}{\textwidth/5-2\tabcolsep-0.2in}
\setlength{\cellWidthe}{\textwidth/5-2\tabcolsep-0.2in}
\scalebox{1}[1]{\begin{tabularx}{\textwidth}{>{\centering\arraybackslash}m{\cellWidtha}>{\centering\arraybackslash}m{\cellWidthb}>{\centering\arraybackslash}m{\cellWidthc}>{\centering\arraybackslash}m{\cellWidthd}>{\centering\arraybackslash}m{\cellWidthe}}
\toprule

\textbf{Model} & \textbf{$\boldmath{-}$2 Log Likelihood} & \textbf{LR $\bm{\upchi}$\textsuperscript{2}} & \textbf{df} & \textbf{\emph{p}}\\
\cmidrule{1-5}

Intercept-only & 63.516 &   &   &  \\

Final & 33.959 & 29.558 & 3 & \textless{}0.001\\

\bottomrule
\end{tabularx}}

    \end{table}

At the category level, PS staff had significantly higher odds of being classified into the institutional barrier category than into the cultural barrier category (\tabref{tabref:education-4286608-t005}). Differences between Ethical and Individual barriers were not statistically significant. This result suggests that PS staff are structurally more likely to frame GenAI constraints in terms of governance, policy clarity, infrastructure, and organizational conditions.    
    \begin{table}[H]
    
    \caption{Job role group effects by selected barrier category (reference outcome = cultural; reference predictor group = academic).}
    \label{tabref:education-4286608-t005}

\begin{adjustwidth}{-\extralength}{0cm}
\setlength{\cellWidtha}{\fulllength/5-2\tabcolsep-0in}
\setlength{\cellWidthb}{\fulllength/5-2\tabcolsep-0.3in}
\setlength{\cellWidthc}{\fulllength/5-2\tabcolsep-0.3in}
\setlength{\cellWidthd}{\fulllength/5-2\tabcolsep-0.3in}
\setlength{\cellWidthe}{\fulllength/5-2\tabcolsep+0.9in}
\scalebox{1}[1]{\begin{tabularx}{\fulllength}{>{\centering\arraybackslash}m{\cellWidtha}>{\centering\arraybackslash}m{\cellWidthb}>{\centering\arraybackslash}m{\cellWidthc}>{\centering\arraybackslash}m{\cellWidthd}>{\centering\arraybackslash}m{\cellWidthe}}
\toprule

\textbf{Outcome Category} & \textbf{B} & \textbf{OR = Exp(B)} & \textbf{\emph{p}} & \textbf{Notes}\\
\cmidrule{1-5}

Ethical & $-$0.196 & 0.822 & 0.522 & Not significant\\

Individual & 0.502 & 1.653 & 0.101 & Not significant \\

Institutional & 0.788 & 2.198 & 0.016 & PS staff higher odds than academics\\

\bottomrule
\end{tabularx}}
\end{adjustwidth}

    \end{table}

Taken together, the MLR analyses provide qualified support for H1. Both discipline and institutional roles are significantly associated with barrier-category distributions, indicating that barriers are associated with institutional position. However, the modest explanatory power of the models also indicates that structural factors shape barrier framing without fully determining it. The most robust differentiation emerges around institutional barriers, particularly across professional roles, supporting a cautious interpretation that GenAI adoption in HE is embedded in organizational and governance contexts rather than reducible to individual perceptions alone.

\subsection{SEM and Hypothesis Evaluation \label{sect:sec5dot3-education-4286608}}

SEM was conducted in SPSS Amos 31 using maximum likelihood, with a sample size \emph{n} = 271 (one incomplete response was removed). As illustrated in \fig{fig:education-4286608-f001}, job role (academic = 1, PS = 0) and discipline (STEM = 1, non-STEM = 0) were specified as exogenous observed predictors. Four barrier categories were modelled as distinct endogenous variables. Guidance, support, attitude, literacy, and job threat were modelled as intermediate endogenous variables (mediating pathways) linking role/discipline to barrier outcomes.

The overall model fit was mixed and is therefore interpreted cautiously for an exploratory path model, indicated by having $\upchi$\textsuperscript{2} = 31.618 with df = 9 (\emph{p} \textless{} 0.001), yielding $\upchi$\textsuperscript{2}/df = 3.513. Incremental fit indices were high (CFI = 0.970, IFI = 0.972, NFI = 0.961), and absolute misfit was low (RMR = 0.034, GFI = 0.980). However, RMSEA was 0.096 (90\% CI [0.061, 0.134], PCLOSE = 0.017). Given the low degrees of freedom, the strong CFI/IFI/NFI values indicate that the model captures the covariance structure well. Taken together, the elevated RMSEA and significant PCLOSE suggest that the SEM results should be interpreted as an exploratory path analysis that identifies selected associations rather than a well-fitted causal model.

SEM provides pathway-level insight that helps interpret H1 and complements the MLR results. The findings support the claim that barrier framing is associated with both disciplinary context and institutional role, beyond individual-level perceptions alone, consistent with H1. At the same time, significant effects were concentrated in selected pathways rather than operating uniformly across the model, as shown in the significant paths listed below:\begin{enumerate}[label=$\bullet$]

\item Role $\rightarrow$ Guidance: b = 0.359, SE = 0.127, CR = 2.833, \emph{p} = 0.005
\item Role $\rightarrow$ Attitude: b = $-$0.498, SE = 0.139, CR = $-$3.594, \emph{p} \textless{} 0.001
\item Role $\rightarrow$ Ethical barriers: b = 0.372, SE = 0.098, CR = 3.786, \emph{p} \textless{} 0.001
\item Role $\rightarrow$ Literacy: b = 0.266, SE = 0.104, CR = 2.549, \emph{p} = 0.011
\item Role $\rightarrow$ Institutional barriers: b = $-$0.215, SE = 0.078, CR = $-$2.763, \emph{p} = 0.006
\item Discipline $\rightarrow$ Attitude: b = 0.579, SE = 0.156, CR = 3.722, \emph{p} \textless{} 0.001
\item Discipline $\rightarrow$ Cultural barriers: b = $-$0.121, SE = 0.057, CR = $-$2.145, \emph{p} = 0.032
\item Attitude $\rightarrow$ Ethical barriers: b = $-$0.179, SE = 0.043, CR = $-$4.200, \emph{p} \textless{} 0.001
\item Attitude $\rightarrow$ Institutional barriers: b = 0.189, SE = 0.034, CR = 5.575, \emph{p} \textless{} 0.001
\item Literacy $\rightarrow$ Individual barriers: b = $-$0.260, SE = 0.055, CR = $-$4.706, \emph{p} \textless{} 0.001
\item Job Threat $\rightarrow$ Ethical barriers: b = 0.138, SE = 0.040, CR = 3.483, \emph{p} \textless{} 0.001

\end{enumerate}

\begin{figure}[H]
      \includegraphics[scale=1]{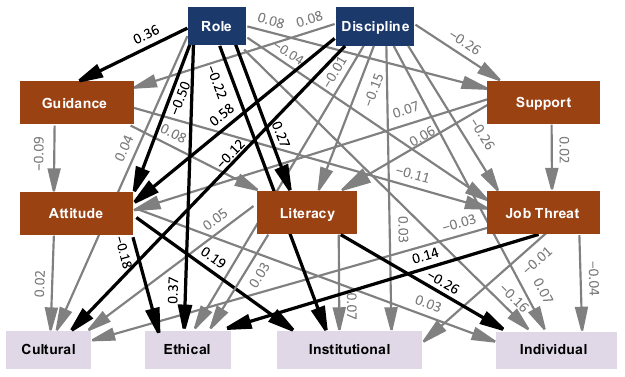}
\caption{SEM model structure and path coefficients, where grey indicates non-significant paths and black indicates significant paths.}
\label{fig:education-4286608-f003}
\end{figure}

Also illustrated in \fig{fig:education-4286608-f003}, discipline (STEM vs. non-STEM) significantly predicted attitude, while role also significantly predicted attitude. Attitude in turn significantly predicted institutional barriers and ethical barriers, indicating that attitudinal differences are linked to both institutional and ethical barrier framing. Role-based differences (academic vs. PS staff) also directly predicted institutional barriers, ethical barriers, guidance, and literacy, while discipline directly predicted cultural barriers. In addition, literacy significantly predicted individual barriers, and job threat significantly predicted ethical barriers.    

SEM also shows that several expected paths were not supported. Although not all pathways were tested or statistically significant in the MLR analyses, the significant SEM paths provide complementary pathway-level evidence that is broadly consistent with H1a and H1b. However, the non-significant paths and cross-sectional design mean that mediation should be interpreted as tentative rather than causal. Although role significantly predicted guidance, guidance and support did not significantly predict attitude, literacy, or job threat. However, role significantly predicted literacy, which in turn significantly predicted individual barriers, indicating an additional indirect role-related pathway. Role also directly predicted ethical barriers, and discipline directly predicted cultural barriers.

\subsection{Free-Text Cluster Themes \label{sect:sec5dot4-education-4286608}}

As described previously in \sect{sect:sec4-education-4286608} Methodology, we first embedded the free-text entries using OpenAI’s text embedding model to represent responses in a 3072-dimensional semantic vector space. We then applied dimensionality reduction using UMAP, followed by HDBSCAN clustering. As reported, clustering settings were selected to balance response coverage, the number of clusters, and the distribution of cluster sizes while still allowing smaller recurring themes to emerge. After forming the clusters, we labelled them using an LLM (gpt-5-mini), and the results are shown in \tabref{tabref:education-4286608-t006}.    
    \begin{table}[H]
    
    \caption{Top 10 barrier clusters from free-text entries.}
    \label{tabref:education-4286608-t006}

\begin{adjustwidth}{-\extralength}{0cm}
\setlength{\cellWidtha}{\fulllength/4-2\tabcolsep-1.4in}
\setlength{\cellWidthb}{\fulllength/4-2\tabcolsep-1.4in}
\setlength{\cellWidthc}{\fulllength/4-2\tabcolsep+0.3in}
\setlength{\cellWidthd}{\fulllength/4-2\tabcolsep+2.5in}
\scalebox{1}[1]{\begin{tabularx}{\fulllength}{>{\centering\arraybackslash}m{\cellWidtha}>{\centering\arraybackslash}m{\cellWidthb}>{\raggedright\arraybackslash}m{\cellWidthc}>{\raggedright\arraybackslash}m{\cellWidthd}}
\toprule

\textbf{ID} & \textbf{Size} & \multicolumn{1}{>{\centering\arraybackslash}m{\cellWidthc}}{\textbf{Description}} & \multicolumn{1}{>{\centering\arraybackslash}m{\cellWidthd}}{\textbf{Representative Phrases}}\\
\cmidrule{1-4}

B1 & 21 & Opposed to GenAI integration & ‘I will not be integrating GenAI into my teaching’, ‘I do not intend to integrate GenAI’, ‘I am strongly against the integration of GenAI’, ‘No, I don’t plan to integrate AI into my teaching’, ‘I have no desire to embed genAI and support a ban’\\
\cmidrule{1-4}
B2 & 20 & Copilot-only policy frustration & ‘Copilot is the only authorised tool’, ‘Copilot is inferior to ChatGPT/Claude’, ‘No approval for competitor products’, ‘Shifting guidance on data uploads’, ‘Poor quality and hallucinations’\\
\cmidrule{1-4}
B3 & 15 & Erodes learning and integrity & ‘Undermines critical thinking and writing’, ‘Students bypass skills for instant gratification’, ‘Makes cheating hard to detect; devalues degrees’, ‘Produces superficially competent but inaccurate work’, ‘Shortcuts before learning underlying skills’\\
\cmidrule{1-4}
B4 & 14 & Eroding critical thinking & ‘Students becoming too reliant on GenAI’, ‘GenAI obliterating students’ critical thinking skills’, ‘Students skip work trusting AI’, ‘Over-reliance on prompting instead of thinking’, ‘GenAI spreading misinformation as fact’\\
\cmidrule{1-4}
B5 & 13 & Factual accuracy and reliability & ‘AI isn’t always correct!’, ‘When AI gets it wrong.’, ‘factual accuracy in finance’, ‘still checking the work’, ‘misinterpreted core message’\\
\cmidrule{1-4}
B6 & 11 & Accuracy and trust concerns & ‘Some staff are mistrustful of it.’, ‘Not being accurate, so always have to check over what the output is.’, ‘It is not personal, and content may end up all sounding the same.’, ‘I don’t trust it to be either ethical or good enough quality.’, ‘Whether the information delivered is factual.’\\
\cmidrule{1-4}
B7 & 11 & Institutional access and support~barriers & ‘No access to latest AI tools-subscriptions required’, ‘University won’t provide licenses or premium versions’, ‘Policies make tool approval time-consuming’, ‘Lack of funding or licencing for AI tools’, ‘Departmental reluctance and low institutional uptake’\\
\cmidrule{1-4}
B8 & 9 & Skill erosion and assessment~validity & ‘Undermines critical thinking’, ‘Students over-rely on AI’, ‘Written essays become invalid’, ‘Homogenized writing styles’, ‘AI-generated factual errors’\\
\cmidrule{1-4}
B9 & 9 & Concerns about AI use & ‘AI encourages shortcut assignments’, ‘Leads to shallow engagement with tasks’, ‘Students distrust teachers using AI’, ‘Teacher dependence may reduce attendance’, ‘Variability, cost and access issues’, ‘Students resent AI-driven grading’\\
\cmidrule{1-4}
B10 & 9 & Unclear policy and guidance & ‘Unclear whether university policy allows it’, ‘No clear guidance on GDPR and confidentiality’, “Don’t know who to consult (IDG/legal)”, ‘Avoid using it due to policy uncertainty’, ‘Difficulty finding policies on university website’\\

\bottomrule
\end{tabularx}}
\end{adjustwidth}

    \end{table}

From \tabref{tabref:education-4286608-t006}, it is evident that free-text clustering broadly complements the multi-level structure identified in the quantitative analysis. Three broad thematic domains \mbox{are observable}.

First, a small but distinct group of responses reflects normative resistance (B1), where staff explicitly reject GenAI integration. These statements are not framed in technical or training terms but express principled opposition, aligning with ethical and cultural \mbox{barrier categories.}

Second, several clusters reflect epistemic and pedagogical risk (B3, B4, B8, B9). Respondents emphasize erosion of critical thinking, shortcut learning, declining academic standards, and threats to assessment validity. This language mirrors the ethical barrier category identified in the survey and is consistent with the descriptive finding that academics, particularly in non-STEM disciplines, foreground integrity and \mbox{skill-development concerns.}

Third, multiple clusters point to institutional governance and infrastructure constraints (B2, B7, B10), including licencing restrictions, unclear policy guidance, GDPR uncertainty, and limited access to tools. These themes correspond directly to the institutional barrier category and align with the multinomial finding that PS staff are more likely to frame barriers organizationally.

Finally, clusters focused on accuracy and trust (B5, B6) cut across levels, indicating persistent concerns about reliability and output quality.

Taken together, the qualitative data supports and contextualizes the role and discipline-related patterns identified quantitatively: while GenAI use is widespread, the perceived barriers are shaped by disciplinary norms, professional role, and governance context rather than by individual reluctance alone.

\section{Discussion \label{sect:sec6-education-4286608}}

\subsection{Adoption Beyond the Individual Level \label{sect:sec6dot1-education-4286608}}

While perceived usefulness and the ease of use remain relevant, they do not fully capture how adoption unfolds in HE, where technologies are embedded within assessment regimes, governance frameworks, professional norms, and regulatory constraints. The findings support growing arguments that technology adoption in AI-enabled environments cannot be understood through individual-level acceptance models alone \mbox{(\citealp{B27-education-4286608})}.

Across analytical stages, a consistent pattern emerges. The descriptive results show that barriers cluster in structured ways rather than appearing as evenly distributed individual concerns. The MLR results confirm that barrier categories are systematically associated with institutional position, even if the explanatory power of single structural variables remains modest. The SEM extends this insight by demonstrating differentiated pathways: literacy strongly predicts capability-based barriers, ethical concerns are linked to perceptions of professional vulnerability, and institutional barriers are associated with governance and organizational conditions. Finally, the text clustering analysis illustrates how these dynamics are expressed in practice, with respondents framing barriers in terms of assessment integrity, policy ambiguity, infrastructure constraints, reliability concerns, and normative resistance.

Taken together, these findings indicate that GenAI adoption in HE emerges from the interaction of user competence, task demands, professional identity, and institutional governance. Individual skills and attitudes matter, but they operate within broader structural conditions that are associated with how usefulness, risk, and responsibility are interpreted. Addressing adoption challenges therefore requires attention not only to training and confidence-building, but also to the institutional and normative environments in which GenAI is deployed.

\subsection{PS Staff as a Distinct Site of GenAI Adoption \label{sect:sec6dot2-education-4286608}}

Another key contribution of the current study is the inclusion of PS staff alongside academics. While recent studies increasingly recognize that GenAI adoption in HE extends beyond teaching and learning, empirical research still tends to focus on classroom contexts and aggregate PS roles into ‘administrator’ categories (\citealp{B37-education-4286608}). This underplays the important functions PS staff perform in enabling institutions to operate, including through governance, compliance, and student services. By systematically comparing academic and PS staff within the same institutional context, we address a significant gap in the literature and provide a structured analysis of both the similarities and differences between the two groups.

Our results suggest that PS staff are more likely to report institutional barriers, and their institutional role has a direct association with this barrier category. We also found that, compared with academics, PS staff also frame the problem differently, emphasizing licencing restrictions, policy ambiguity, and governance processes. This pattern is consistent with emerging evidence from the literature, but our findings extend this work by showing that HE institutions face multiple adoption problems simultaneously: a pedagogical adoption problem in academic domains such as academic integrity and student learning experience and outcomes, and an organizational adoption problem in PS domains where challenges are more strongly linked to governance and implementation challenges embedded in institutional systems and accountability structures.

The findings have practical significance for institutional strategy. First, role-specific implementation guidance is required, particularly around tool approval, procurement processes, and operational use cases. This is important not only for addressing current barriers such as the ‘Copilot-only policy’ (e.g., B2 in \tabref{tabref:education-4286608-t006}), but also for helping institutions remain adaptive in a rapidly evolving GenAI ecosystem. Second, our results support recent calls to widen GenAI adoption research beyond pedagogy and to examine the full institutional ecology. In this respect, PS staff should be treated as a distinct analytical group whose experiences are essential for understanding GenAI adoption in HE.

\subsection{Gains and Trade-Offs of the Multi-Method Design \label{sect:sec6dot3-education-4286608}}

The current study employs a multi-method analytical design based on the data collected from a staff survey. This allows us to move beyond simple descriptive accounts and toward a more structured explanation and comparison. Recent progress in the literature highlights existing work in GenAI adoption (\citealp{B11-education-4286608}; \citealp{B46-education-4286608}; \mbox{\citealp{B49-education-4286608}}); however, descriptive statistics or single-model acceptance frameworks often fail to address relational dynamics through which adoption unfolds in complex institutional settings. Our analytical pipeline takes advantage of and combines several approaches: descriptive analyses establish empirical patterns; MLR and SEM test statistical significance and estimate pathway-level relationships among roles, disciplines, and mediating variables; and the clustering of free-text responses captures themes and provides context that is difficult to fully represent in predefined survey items. For example, the MLR showed that PS staff had higher odds of selecting institutional barriers than academics; this is echoed in the clustering results, where respondents referred to Copilot-only restrictions, licencing limitations, and unclear policy guidance. Similarly, the MLR findings showed that non-STEM academics were more strongly associated with ethical and pedagogical concerns; this is reflected in clusters that centre on academic integrity, over-reliance, erosion of critical thinking, and assessment validity. Taken together, this multi-method design provides a richer account of GenAI adoption than would be possible through any \mbox{single approach.}

However, there are certain caveats.

First, MLR showed modest pseudo-\emph{R}\textsuperscript{2} values, indicating that role and discipline are meaningful but incomplete predictors of barrier-category membership. The remaining heterogeneity may reflect factors beyond the survey, including seniority, teaching intensity, and departmental culture. The models’ modest explanatory power, along with single-institutional design and non-probability sampling, therefore point to the need for more granular accounts of institutional and professional contexts in the future.

Second, SEM improves purely descriptive analyses; however, causal claims such as from roles to different barriers should be made cautiously. At present, the evidence supports associations rather than causal effects. Third, free-text clustering and LLM-assisted labelling provide scalable thematic analysis, but they involve parameter fine-tuning that can substantially influence cluster granularity and interpretation (\citealp{B4-education-4286608}; \citealp{B43-education-4286608}). Finally, a further limitation concerns the use of STEM versus non-STEM as the primary disciplinary variable. This binary offers a useful proxy for broad disciplinary differences, but it collapses substantial within-group variation across fields with different methodological traditions, assessment formats, uses of GenAI, and risk perceptions. We therefore interpret these findings cautiously. A finer-grained taxonomy was considered but not adopted because several subgroup sizes were too small for reliable multinomial modelling or SEM estimation. Future studies with larger, multi-institutional samples should disaggregate disciplinary contexts more precisely.

\section{Conclusions \label{sect:sec7-education-4286608}}

This study investigated the barriers to GenAI adoption at different levels through a multi-method survey of 272 academic and PS staff at Russell Group university. By integrating MLR, SEM, and semantic clustering, we found evidence that reported that adoption barriers are not merely individual but are associated with institutional roles and disciplinary contexts. Our findings show that non-STEM academics primarily face ethical and pedagogical concerns, whereas STEM and PS staff disproportionately encounter institutional and governance constraints.

This research makes three contributions with significant implications for university policy and technology adoption frameworks. First, by providing a systematic cross-disciplinary comparison within a single-institutional environment, we highlight how distinct epistemic norms dictate the risks of GenAI, suggesting that generic, university-wide adoption policies are likely to be ineffective. Non-STEM academics may need support focused on assessment integrity, authorship, disciplinary standards, and critical thinking. STEM academics may benefit from applied guidance on reliability, verification, reproducibility, and technical workflow integration. PS staff require clearer operational guidance on approved tools, data protection, licencing, procurement, confidentiality, and governance escalation routes. Institutional responses should therefore combine general GenAI literacy with differentiated guidance documents, role-specific training, discipline-sensitive assessment resources, and transparent mechanisms for tool approval and responsible use.

Second, extending the analytical focus beyond teaching academics to include PS staff uncovers a critical yet frequently overlooked dimension of organizational adoption centred on compliance, licencing, and student services. Third, our methodological integration of quantitative modelling with LLM-assisted qualitative clustering advances the field past descriptive analyses toward multi-level, structural explanations. Consequently, HE institutions can leverage these insights to design targeted, role-specific governance processes, frameworks, and strategies that directly address the localized adoption barriers for their diverse workforce.

In addition, this study presents opportunities for future research. The data rely on a cross-sectional, non-probability sample from a single institution. While SEM highlighted key associations, it cannot definitively establish causal relationships, for example, between institutional positions and specific adoption barriers. Additionally, the modest explanatory power of our regression models suggests that other unmeasured variables such as departmental culture may also play pivotal roles. Future longitudinal studies across diverse institutional types, coupled with in-depth qualitative interviews, will be necessary to validate these structural pathways and track how GenAI barrier perceptions evolve as institutional policies mature.

\vspace{6pt}
\authorcontributions{Conceptualization, J.Y. and K.Ö.; methodology, J.Y., K.Ö., A.v.M. and A.B.A.; software, J.Y.; writing---original draft preparation, J.Y., K.Ö. and A.B.A.; writing---review and editing, A.v.M., T.S.C. and C.O.; funding acquisition, J.Y. All authors have read and agreed to the published version of the manuscript.}
\funding{This research was funded by University of Warwick Education Fund.}
\institutionalreview{The study was conducted in accordance with the Declaration of Helsinki and approved by Warwick’s Biomedical and Scientific Research Ethics Committee (BSREC) (reference number BSREC 85/24-25 and date of approval: 25 February 2025).}
\informedconsent{Informed consent was obtained from all subjects involved in the~study.}
\dataavailability{The data that support the findings of this study are available from the corresponding author upon reasonable request, subject to non-disclosure requirements.}
\acknowledgments{The authors wish to thank Warwick Education Fund (project “Identifying Barriers and Use Cases for Generative AI in Education Using Retrieval-Augmented Generation and Staff Surveys”) for financial support.

}
\conflictsofinterest{The authors disclose that they have no actual or perceived conflicts of interest.}
\abbreviations{Abbreviations}{~The following abbreviations are used in this manuscript:\\

\noindent
\begin{tabular}{@{}ll}

AI & Artificial Intelligence\\

GenAI & Generative Artificial Intelligence\\

GDPR & General Data Protection Regulation\\

HDBSCAN & Hierarchical Density-Based Spatial Clustering of Applications with Noise\\

HE & Higher Education\\

LLM & Large Language Model\\

MLR & Multinomial Logistic Regression\\

PS & Professional Services\\

SEM & Structural Equation Modelling\\

STEM & Science, Technology, Engineering, and Mathematics\\

TAM & Technology Acceptance Model\\

UMAP & Uniform Manifold Approximation and Projection\\

UTAUT & Unified Theory of Acceptance and Use of Technology\\

\end{tabular}
}
\begin{adjustwidth}{-\extralength}{0cm}

\reftitle{References}

\end{adjustwidth}
\begin{adjustwidth}{-\extralength}{0cm}
\PublishersNote{}
\end{adjustwidth}

\end{document}